\documentclass[times, 10pt,twocolumn]{article}
\usepackage{latex8}
\usepackage{times}

\usepackage{graphicx}
\usepackage{epsfig}
\usepackage{amsmath}
\usepackage{amssymb}

\twocolumn

\begin{document}

\newcommand{\ignore}[1]{}
\newcommand{\db}[1]{{\sffamily\bfseries\Large \mbox{***DB:\ } #1 \mbox{****
}}}
\newcommand{\sk}[1]{{\sffamily\bfseries\Large \mbox{***SK:\ } #1 \mbox{****
}}}
\newcommand{\comment}[1]  {}
\newcommand\ie{{\textsl{i.e.\,}}}
\newcommand\eg{{\textsl{e.g.\,}}}
\newcommand\etal{{\textsl{et al.\,}}}
\def\BE{\begin{equation}}
\def\EE{\end{equation}}
\def\BEA{\begin{eqnarray}}
\def\EEA{\end{eqnarray}}
\newcommand{\cut}[1]{{}}
\newcommand\va{{\bf a}} 
\newcommand\vb{{\bf b}}
\newcommand\vc{{\bf c}}
\newcommand\vd{{\bf d}}
\newcommand\ve{{\bf e}}
\newcommand\vf{{\bf f}}
\newcommand\vg{{\bf g}}
\newcommand\vh{{\bf h}}
\newcommand\vi{{\bf i}}
\newcommand\vj{{\bf j}}
\newcommand\vk{{\bf k}}
\newcommand\vl{{\bf l}}
\newcommand\vm{{\bf m}}
\newcommand\vn{{\bf n}}
\newcommand\vo{{\bf o}}
\newcommand\vp{{\bf p}}
\newcommand\vq{{\bf q}}
\newcommand\vr{{\bf r}}
\newcommand\vs{{\bf s}}
\newcommand\vt{{\bf t}}
\newcommand\vu{{\bf u}}
\newcommand\vv{{\bf v}}
\newcommand\vw{{\bf w}}
\newcommand\vx{{\bf x}}
\newcommand\vy{{\bf y}}
\newcommand\vz{{\bf z}}
\newcommand\mA{{\bf A}} 
\newcommand\mB{{\bf B}}
\newcommand\mC{{\bf C}}
\newcommand\mD{{\bf D}}
\newcommand\mE{{\bf E}}
\newcommand\mF{{\bf F}}
\newcommand\mG{{\bf G}}
\newcommand\mH{{\bf H}}
\newcommand\mI{{\bf I}}
\newcommand\mJ{{\bf J}}
\newcommand\mK{{\bf K}}
\newcommand\mL{{\bf L}}
\newcommand\mM{{\bf M}}
\newcommand\mN{{\bf N}}
\newcommand\mO{{\bf O}}
\newcommand\mP{{\bf P}}
\newcommand\mQ{{\bf Q}}
\newcommand\mR{{\bf R}}
\newcommand\mS{{\bf S}}
\newcommand\mT{{\bf T}}
\newcommand\mU{{\bf U}}
\newcommand\mV{{\bf V}}
\newcommand\mW{{\bf W}}
\newcommand\mX{{\bf X}}
\newcommand\mY{{\bf Y}}
\newcommand\mZ{{\bf Z}}

\title{Peer-to-Peer Secure Multi-Party Numerical Computation}
\author{Danny Bickson\thanks{The work on this paper was done when DB was a Ph.D. student at the Hebrew University of Jerusalem. Supported by The Israel Science Foundation
  (grant No.~0397373).}\\
IBM Haifa Research Lab,\\Mount Carmel, Haifa 31905, Israel.\\ dannybi@il.ibm.com\\
\and %
Danny Dolev, Genia Bezman\\
School of Computer Science and Engineering\\
The Hebrew University of Jerusalem\\
Jerusalem 91904, Israel.\\
dolev,genia4@cs.huji.ac.il\\
\and Benny Pinkas\thanks{Supported by The Israel Science Foundation
  (grant No.~860/06).}\\
Dept. of Computer Science,\\ University of Haifa, Mount Carmel, Haifa 31905, Israel.\\
benny@pinkas.net\\
}

\maketitle
\thispagestyle{empty}

\begin{abstract}
We propose an efficient framework for enabling secure multi-party
numerical computations in a Peer-to-Peer network. This problem
arises in a range of applications such as collaborative filtering,
distributed computation of trust and reputation, monitoring and
numerous other tasks, where the computing nodes would like to
preserve  the privacy of their inputs while performing  a joint
computation of a certain function.

Although there is a rich literature in the field of distributed
systems security concerning secure multi-party computation, in
practice it is hard to deploy those methods in very large scale
Peer-to-Peer networks. In this work, we examine several possible
approaches and discuss their feasibility. Among the possible
approaches, we identify a single approach which is both scalable
and theoretically secure.

An additional novel contribution is that we show how to compute the
neighborhood based collaborative filtering, a state-of-the-art
collaborative filtering algorithm, winner of the Netflix progress
prize of the year 2007. Our solution computes this algorithm in a
Peer-to-Peer network, using a privacy preserving computation,
without loss of accuracy.

Using extensive large scale simulations on top of real Internet
topologies, we demonstrate the applicability of our approach. As
far as we know, we are the first to implement such a large scale
secure multi-party simulation of networks of millions of nodes
and hundreds of millions of edges.
\end{abstract}

\section{Introduction}
We consider the problem of performing a joint numerical
computation of some function over a Peer-to-Peer network. Such
problems arise in many applications, for example, when computing
distributively trust~\cite{EigenTrust}, ranking of nodes and data
items~\cite{p2p-rating}, clustering~\cite{EWSN08}, collaborative
filtering~\cite{KorenCF,PP2}, factor analysis~\cite{Canny} etc.
The aim of {\em secure multi-party computation} is to enable
parties to carry out such distributed computing tasks in a secure
manner. Whereas distributed computing classically deals with
questions of computing under the threat of machine crashes and
other inadvertent faults, secure multi-party computation is
concerned with the possibility of deliberate malicious behavior by
some adversarial entity. That is, it is assumed that a protocol
execution may come under attack by an external entity, or even by
a subset of the participating parties. The aim of this attack may
be to learn private information or cause the result of the
computation to be incorrect. Thus, two central requirements on any
secure computation protocol are privacy and correctness. The
privacy requirement states that nothing should be learned beyond
what is absolutely necessary; more exactly, parties should learn
their designated output and nothing else. The correctness
requirement states that each party should receive its correct
output. Therefore, the adversary must not be able to cause the
result of the computation to deviate from the function that the
parties had set out to compute.

\ignore{ More formally, the security of a protocol is established
by comparing the outcome of a real protocol execution to the
outcome of an ideal computation. That is, for any adversary
attacking a real protocol execution, there exists an adversary
attacking an ideal execution (with a trusted party) such that the
input/output distributions of the adversary and the participating
parties in the real and ideal executions are essentially the same.
Thus a real protocol execution emulates the ideal world. This
formulation of security is called the ideal/real simulation
paradigm. Privacy follows from the fact that the adversary's
output is the same in the real and ideal executions. Since the
adversary learns nothing beyond the corrupted party's outputs in
an ideal execution, the same must be true for a real execution.
Correctness follows from the fact that the honest parties' outputs
are the same in the real and ideal executions, and from the fact
that in an ideal execution, the honest parties all receive correct
outputs as computed by the trusted party.}

In this paper, we consider only functions which are built using
the algebraic primitives of addition, substraction and
multiplication. In particular, we  focus on numerical methods
which are computed distributively in a Peer-to-Peer network, where
in each iteration, every node interacts with a subset of its
neighbors by sending scalar messages, and computing a weighted sum
of the messages that it receives. Examples of such functions are
belief propagation~\cite{BibDB:BookPearl}, EM (expectation
maximization) ~\cite{Canny}, Power method~\cite{EigenTrust},
separable functions~\cite{Separable}, gradient descent
methods~\cite{PP3} and linear iterative algorithms for solving
systems of linear equations~\cite{BibDB:BookBertsekasTsitsiklis}.
As a specific example, we describe the Jacobi algorithm in detail
in Section~\ref{Jacobi}.

There is a rich body of research on secure computation, starting
with the seminal work of Yao~\cite{Yao}. Part of this research is
concerned with the design of {\em generic} secure protocols that
can be used for computing any function (for example, Yao's
work~\cite{Yao} for the case of two participants, and
e.g.~\cite{BGW,GMW} for solutions for the case of multiple
participants).  There are several works concerning the {\em
implementation} of generic protocols for secure computation. For
example, FairPlay~\cite{Fairplay} is a system for secure two-party
computation, and FairPlayMP~\cite{FairPlayMP} is a different
system for secure computation by more than two parties.  These two
systems are based (like Yao's protocol) on reducing any function
to a representation as a Boolean circuit and computing the
resulting Boolean circuit securely. Our approach is much more
efficient, at the cost of supporting only a subset of the
functions the FairPlay system can compute.

A different line of work studies secure protocols for computing
specific functions (rather than generic protocols for computing any
function). Of particular interest for us are works that add
a privacy preserving layer to the computation of functions such as the
factor analysis learning problem (for which~\cite{Canny} describes a
secure multi-party protocol using homomorphic encryption),
computing trust in a Peer-to-Peer network (for which~\cite{EigenTrust}
suggests a solution using a trusted third party), or the work
of~\cite{PP3}, which is closely related to our work, but is limited to
two parties.

Most  previous solutions for secure multi-party computation suffer
from one of the following drawbacks: (1) they provide a
centralized solution where all information is shipped to a single
computing node, and/or (2) require communication between all
participants in the protocol, and/or (3) require the use of asymmetric
encryption, which is costly. In this work, we investigate
 secure computation in a Peer-to-Peer setting, where each
node is only connected to some of the other nodes (its neighbors).
We examine different possible approaches, and out of the different
approaches we identify a single approach, which is theoretically
secure, efficient, and scalable.

Security is often based on the assumption that there is an upper
bound on the {\em global} number of malicious participants. In our
setting, we consider the number of malicious nodes in each {\em
local vicinity}. Furthermore, most of the existing algorithms
scale to tens or hundreds of nodes at the most. In this work, we address
the problem in a setting of a large Peer-to-Peer network, with
millions of nodes and hundreds of millions of communication links.
Unlike most of the previous work, we have performed a {\em very
large scale simulation}, using real Internet topologies, to show
our approach is applicable to real network settings.

As an example for applications of our framework, we take the
neighborhood based collaborative filtering~\cite{KorenCF}. This
algorithm is a recent state-of-the-art algorithm. There are two
challenges in adapting this algorithm to a Peer-to-Peer network.
First, the algorithm is centralized and we propose a method to
distribute it. Second, we add a privacy preserving layer, so no
information about personal ranking is revealed during the process
of computation.

 The paper is organized as follows. In Section~\ref{model} we
formulate our problem model. In Section~\ref{crypto} we give a
brief background of cryptographic primitives that are used in our
schemes. Section~\ref{const} outlines our novel construction. We
give a detailed case study of collaborative filtering as an
example application in Section~\ref{jacobi}. Large scale
simulations are presented in Section~\ref{exp}. We conclude in
Section~\ref{Conclusion}.

We use the following notations: $T$ stands for a vector or matrix
transpose, the symbols $\{\cdot\}_{i}$ and $\{\cdot\}_{ij}$ denote
entries of a vector and matrix, respectively. The spectral radius
$\rho(\mB)\triangleq\max_{1\leq i\leq s}(|\lambda_{i}|)$, where
$\lambda_{1},\ldots\lambda_{s}$ are the eigenvalues of a matrix
$\mB$. $N_i$ is the set of neighboring nodes to node $i$.

\section{Our Model}
\label{model} Given a Peer-to-Peer network graph $G=(V, E)$ with
$|V| = n$ nodes and $|E| = e$ edges, we would like to perform a
joint iterative computation. Each node $i$ starts with a scalar
state\footnote{An extension to the vector case is immediate, we
omit it for the clarify of description.} $x_i^0 \in \mathbb{R}$,
and on each round, sends messages to a subset of its neighbors. We
denote a message sent from node $i$ to node $j$ at round $r$ as
$m_{i,j}^r$.

Let $N_i$ denote the set of neighboring nodes of $i$.  Denote the
neighbors of node $i$ as $n_{i_1},n_{i_2},\ldots,n_{i_k}$, where
$k=|N_i|$.  We assume, wlog, that each node sends a message to each of
its neighbors.  On each round $r = 1,2,\cdots$, node $i$ computes,
based on the messages it received, a function $f: R^{k+1} \rightarrow
R^{k+1}$, \[ \langle x_i^r, m^{r}_{i,n_{i_1}}, \cdots, m^{r}_{i,n_{i_k}}\rangle =
f(x_i^{r-1}, m^{r-1}_{n_{i_1},i}, \cdots , m^{r-1}_{n_{i_k},i}) \]
Namely, the function gets as input the initial state and all the
received neighbor messages of this round and outputs a new state and
messages to be sent to a subset of the neighbors at the next
round. The iterative algorithms are run either a predetermined number
of rounds, or until convergence is detected locally.

In this paper, we are only interested in functions $f$ which
compute weighted sums on each iteration. Next we show that there
is a variety of such numerical methods. Our goal is to add a
privacy preserving layer to the distributed computation, such that
the only information learned by a node is its share of the output.


We use the semi-honest adversaries model: in this model (common in
cryptographic research of secure computation) even corrupted parties
are assumed to correctly follow the protocol specification. However,
the adversary obtains the internal states of all the corrupted parties
(including the transcript of all the messages received), and attempts
to use this information to learn information that should remain
private.\footnote{Security against semi-honest adversaries might be
  justified if the parties participating in the protocol are somewhat
  trusted, or if we trust the participating parties at the time they
  execute the protocol, but suspect that at a later time an adversary
  might corrupt them and get hold of the transcript of the information
  received in the protocol.

  We note that protocols secure against malicious adversaries are
  considerably more costly than their semi-honest counterparts. For
  example, the generic method of obtaining security against malicious
  adversaries is through the GMW compiler~\cite{GMW} which adds a
  zero-knowledge proof for every step of the protocol.}  In
Section~\ref{Conclusion} discuss the possibility for extending our
construction to the ``malicious adversary'', which can behave
arbitrarily.


We define a configurable local system parameter $d_i$, which
defines the maximum number of nodes in the local vicinity of node $i$
(direct neighbors of node $i$) which are corrupt. Whenever this
assertion is violated, the security of our proposed scheme is
affected. This is a stronger requirement from our system, relative
to the traditional global bound on the number of adversarial
nodes.
\section{Cryptographic primitives}
\label{crypto} We compare several existing approaches from the
literature of secure multi-party computation and discuss their
relevance to Peer-to-Peer networks.

\subsection{Random perturbations}
The random additive perturbation method attempts to preserve the
privacy of the data by modifying values of the sensitive
attributes using a randomized process
(see~\cite{AS,DiNi,RandomNoise}). In this approach, the node sends
a value $u_i + v$, where $u_i$ is the original scalar message, and
$v$ is a random value drawn from a certain distribution $V$. In
order to perturb the data, $n$ independent samples $v_1, v_2,
\cdots , v_n$, are drawn from a distribution $V$. The owners of
the data provide the perturbed values $u_1+v_1, u_2+v_2, \cdots ,
u_n+v_n$ and the cumulative distribution function $FV(r)$ of $V$.
The goal is to use these values, instead of the original ones, in
the computation. (It is easy to see, for example, that if the
expected value of $V$ is $0$, then the expectation of the sum of
the $u_i+v_i$ values is equal to the expectation of the $u_i$
values.) The hope is that by adding random noise to the individual
data points it is possible to hide the individual values.

\ignore{ The $n$ original data values $u_1, u_2, \cdots , u_n$ are
  viewed as realizations of $n$ independent and identically
  distributed (i.i.d.) random variables $U_i, i = 1, 2, \cdots , n$,
  each with the same distribution as that of a random variable U.  In
  order to perturb the data, $n$ independent samples $v_1, v_2, \cdots
  , v_n$, are drawn from a distribution $V$ . The owner of the data
  provides the perturbed values $u_1+v_1, u_2+v_2, \cdots , u_n+v_n$
  and the cumulative distribution function $FV(r)$ of $V$. The
  reconstruction problem is to estimate the distribution $FU(x)$ of
  the original data, from the perturbed data.
}

The random perturbation model is limited. It supports only
addition operations, and it was shown in~\cite{DiNi} that this
approach can ensure very limited privacy guarantees. We only
demonstrate this method as a lightweight protocol, mainly for
comparing its running time with the other protocols.

\subsection{Shamir's Secret Sharing (SSS)}
Secret sharing is a fundamental primitive of cryptographic
protocols. We will describe the  secret sharing scheme of
Shamir~\cite{SSS}. The scheme works over a field $F$, and we
assume the secret $s$ to be an element in that field. In a
$k$-out-of-$n$ secret sharing the owner of secret wishes to
distribute it between $n$ players such that any subset of $k$ of
them is able to recover the secret, while no subset of $k-1$
players is able to learn any information about the secret.

In order to distribute the secret, its owner chooses a random
polynomial $P()$ of degree $k-1$, subject to the constraint that
$P(0)=s$. This is done by choosing random coefficients
$a_1,\ldots,a_{k-1}$ and defining the polynomial as
$P(x)=s+\sum_{i=1}^{k-1}a_ix^i$. Each player is associated with an
identity in the field (denoted $x_1,\ldots,x_n$\/ for players
$1,\ldots,n$, respectively). The share that player $i$ receives is the
value $P(x_i)$, namely the value of the polynomial evaluated at the
point $x_i$.  It is easy to see that any $k$ players can recover the
secret, since they have $k$ values of the polynomial and can therefore
interpolate it and compute its free coefficient $s$. It is also not
hard to see that any set of $k-1$ players does not learn any
information about $s$, since any value of $s$ has a probability of
$1/|F|$ of resulting in a polynomial which agrees with the values that
the players have.
\subsection{Homomorphic encryption}
A homomorphic encryption scheme is an encryption scheme which
allows certain algebraic operations to be carried out on the
encrypted plaintext, by applying an efficient operation to the
corresponding ciphertext (without knowing the decryption key!).
In particular, we will be interested in additively homomorphic
encryption schemes: Here, the message space is a ring (or
a field). There exists an efficient algorithm $+_{pk}$
whose input is the public key of the encryption scheme and two
ciphertexts, and whose output is $E_{pk}(m_1) +_{pk} E_{pk}(m_2) =
E_{pk}(m_1 + m_2)$. (Namely, this algorithm computes, given the
public key and two ciphertexts, the encryption of the sum of the
plaintexts of two ciphertexts.) There is also an efficient
algorithm $\cdot_{pk}$, whose input consists of the public key of
the encryption scheme, a ciphertext, and a constant $c$ in the
ring, and whose output is  $c\cdot_{pk} E_{pk}(m) = E_{pk}(c
\cdot_{pk} m)$.

We will also  require that the encryption scheme has semantic security.
An efficient implementation of an
additive homomorphic encryption scheme with semantic security was
given by Paillier~~\cite{Paillier}. In this cryptosystem the
encryption of a plaintext from $[1;N]$, where N is an RSA modulus,
requires two exponentiations modulo $N^2$. Decryption requires a
single exponentiation. We will use this encryption scheme in our work.

\subsubsection{Paillier encryption}
We describe in a nutshell the Paillier cryptosystem. Fuller
details are found on~\cite{Paillier}.
\begin{itemize}
    \item {\bf Key generation} Generate two large primes p
and q. The secret key $sk$ is $\lambda = lcm(p - 1, q - 1)$. The
public key $pk$ includes $N = pq$ and $g  \in \mathbb{Z}_{N^2}$
such that $g \equiv \mbox{  }1 \mbox{  } \mod N$.
    \item {\bf Encryption} Encrypt a message $m \in \mathbb{Z}_{N}$ with randomness
$r \in \mathbb{Z}_{N^2}^*$ and public key $pk$ as $c = g^mr^N \mod
N^2$.
    \item {\bf Decryption} Decrypt a ciphertext $c \in \mathbb{Z}_{N^2}^*$.
    Decryption is done using: $\frac{L(c^\lambda \mod N^2)}{L(g^\lambda \mod
    N^2)} \mod N$ where $L(x) = (x - 1)/N$.
\end{itemize}

\section{Our construction}
\label{const} The main observation we make is that numerous
distributed numerical methods compute in each node a weighted sum
of scalars $m_{ji}$, received from neighboring nodes, namely \BE
\label{ws} \sum_{j\in N_i} a_{ij} m_{ji}, \EE where the weight
coefficients $a_{ij}$  are known constants. This simple
building block, captures the behavior of multiple numerical
methods. By showing ways to compute this weighted sum securely,
our framework can support many of those numerical methods. In this
section we introduce three possible approaches for performing the
weighted sum computation.

In Section~\ref{Jacobi} we give an example of the Jacobi algorithm
which computes such a weighted sum on each iteration.
\subsection{Random perturbations}
\ignore{ In this method, we only support distributed calculations
of sums. } In each iteration of the algorithm, whenever a node
needs to send a value $m_{ji}$ to a neighboring node, the node
generates a random number $r_{j,i}$ using the GMP
library~\cite{GMP}, from a probability distribution with zero
mean. It then sends the value $m_{ji}+r_{j,i}$ to the other node.
 As the number of neighbors increases,
the computed noisy sum $\sum_{j\in N_i} (m_{ji}+r_{j,i})$
converges to the actual sum $\sum_{j\in N_i} m_{ji}$.

When the node computes a weighted sum of the messages it received
as in equation~\ref{ws}, it multiplies each incoming message by
the corresponding weight. The computed noisy sum $\sum_{j\in N_i}
a_{ij}(m_{ji}+r_{j,i})$
 converges to the
actual sum  $\sum_{j\in N_i} a_im_{ji}$.

We note again that this method is considered mainly for a comparison
of its  running
time with that of  the other methods.

\subsection{Homomorphic Encryption}
%

We chose to utilize the Paillier encryption scheme, which is an
efficient realization of an additive homomorphic encryption scheme
with
semantic security.\\
\\
{\bf Key generation: } We use the threshold version of the
Paillier encryption scheme described in~\cite{Paillier2}. In this
scheme, a trusted third party generates for each node $i$ private
and public key pairs.\footnote{It is also possible to generate the key
  in a distributed way, without using any trusted  party. This option
  is less efficient. We  show that even the usage of a
  centralized key generation process is not efficient enough, and
  therefore we did not implement the distributed version of this protocol.}
  The public key is disseminated to all of
node $i$ neighbors. The private key $\lambda_i=prvk(i)$ is kept
secret from all nodes (including node $i$). Instead, it is split,
using secret sharing, to the neighbors of node $i$. There is a
threshold $d_i$, which is at most equal to $|N_i|$, the number of
neighbors of node $i$. The scheme ensures that any subset of $d_i$
of the neighbors of node $i$ can help it decrypt messages (without
the neighbors learning the decrypted message, or node $i$ learning
the private key). If $d_i=|N_i|$ then the private key is shared by
giving each neighbor $j$ a random value $s_{ji}$ subject to the
constraint $\sum_{j\in N_i}s_{ji} = \lambda_i=prvk(i)$. Otherwise,
if $d_i<|N_i|$ the values $s_{ji}$ are shares of a Shamir secret
sharing of $\lambda_i$. Note that fewer than $d_i$ neighbors cannot
recover the key.

Using this method, all neighboring nodes of node $i$ can send
encrypted messages using $pubk(i)$ to node $i$, while node $i$ cannot
decrypt any of these messages. It can, however,
aggregate the messages using the homomorphic property and ask a  coalition
of $d_i$ or more neighbors to help it in decrypting the sum.\\

The initialization step of this protocol is as follows:
\begin{itemize}
\item [H0] The third party creates for node $i$ a public and private key
  pair, $[pubk(i), prvk(i)]$. It sends the public key $pubk(i)$ to all
  of node $i$'s neighbors, and splits  the private key into
  shares, such that each node $i$ neighbors gets a share
  $s_{ji}$. If $d_i=|N_i|$ then
 $prvk(i)=\lambda_i = \sum_{j \in N_i} s_{ji}$. Otherwise the $s_{ji}$
 values are Shamir shares of the private key.
\end{itemize}

{\bf One round of computation: } In each round of the algorithm,
when a node $j$ would like to send a scalar value $m_{ji}$ to node
$i$ it does the following:
\begin{itemize}
    \item [H1] Encrypt the message $m_{ji}$, using node $i$ public
    key to get $C_{ji} = E_{pubk(i)}(m_{ji})$.
    \item [H2] Send the result $C_{ji}$ to node $i$.
    \item [H3] Node $i$ aggregates all the incoming message $C_{ji}$, using the homomorphic
    property to get $E_{pubk(i)}(\sum a_{ij}m_{ji})$
\end{itemize}

\noindent {\bf After receiving all messages:} Node $i$'s neighbors
assist it in decrypting the result $x_i$, without revealing the
private key $prvk(i)$. This is done as follows (for the case
$d_i=|N_i|$): Recall that in a Paillier decryption node $i$ needs to
raise the result computed in [H3] to the power of its private key
$\lambda_i$.
\begin{itemize}
    \item [H4] Node $i$ sends all its neighbors the result
    computed in [H3]: $C_i = E_{pubk(i)}(\sum a_{ij}m_{ji})$.
    \item [H5] Each neighbor, computes a part of the
    decryption $w_{ji} = C_i ^{s_{ji}}$ where $s_{ji}$ are node $i$ private
    key shares computed in step [H0], and sends the result
    $w_{ji}$ to node $i$.
    \item [H6] Node $i$ multiplies all the received values to get:
    \BE \Pi_{j \in N_i} w_{ji} = C_i ^{\sum_{j \in N_i} s_i} = C_i
    ^{\lambda_i} = \sum a_{ij} m_{ji} \;\; \mathrm{mod} \;\; N.
\EE
\end{itemize}
If $d_i<|N_i|$ then the reconstruction is done using Lagrange
interpolation in the exponent, where node $i$ needs to raise each
$w_{ji}$ value by the corresponding Lagrange coefficient, and then
multiply the results.

Regarding message overhead, first we need to generate and
disseminate public and private keys. This operation requires $2e$
messages, where $e = |E|$ is the number of graph edges. In each
iteration we send the same number of message as in the original
numerical algorithm. However, assuming a security of $\ell$ bits,
and a working precision of $d$ bits, we increase the size of the
message by a factor of $\frac{ \ell}{d}$. Finally, we add $e$
messages for obtaining the private keys parts in step H4.

Regarding computation overhead, for each sent message, we need to
perform one Paillier encryption in step H1. In step H3 the
destination node performs additional $k-1$ multiplications, and
one decryption in step H4. At the key generation phase, we add
generation of $n$ random polynomial and their evaluation. In step
H4 we compute an extrapolation of those $n$ polynomials.
The security of the Paillier encryption is investigated
in~\cite{Paillier, Paillier2}, where it was shown that the system
provides semantic security.

\subsection{Shamir Secret Sharing}
We propose a  construction based on Shamir's secret sharing,
which avoids the computation cost of asymmetric encryption. In a
nutshell, we use the neighborhood of a node for adding a privacy
preserving mechanism, where only a coalition of $d_i$ or more nodes
can reveal the content of messages sent to that node.

In each round of the algorithm, when a node $j$ would like to send
a scalar value $m_{ji}$ to node $i$ it does the following:
\begin{itemize}
    \item [S1] Generate a random polynomial $P_{ji}$ of degree
    $d_i-1$, of the type $P_{ji}(x) = m_{ji} + \sum_{i=1}^{d_i-1} a_i
    x^i$\/ (where $d_i \le |N_i| $).
    \item [S2] For each neighbor $l$ of node $i$, create a share $C_{jil}$ of
    the polynomial $P_{ji}(x)$ by evaluating it on a single point
    $x_{l}$.
    \item [S3] Send $C_{jil}$ to node $l$, which is $i$'s
    neighbor.
    \item [S4] Each neighbor $l$ of node $i$ aggregates the
      shares it received from all neighbors of node $i$  and computes the value
 $S_{li} = \sum_{j \in N_i} a_{ij} P_{ji}(x_{l})$. (Note
 that the result of this computation is equal to the value of a polynomial of degree $d_i-1$,
 whose free coefficient is equal to the {\em weighted} sum of all messages sent to
 node $i$ by its neighbors.)
    \item [S5] Each neighbor $l$ sends the sum $S_{li}$ to node $i$.
    \item [S6] Node $i$ treats the value received from node $l$ as a
      value of a polynomial of degree $d_i-1$ evaluated at the point
      $x_i$.
\ignore{
aggregates all the sums using the homomorphic
    property to get $P_i(x)$}
    \item [S7] Node $i$ interpolates $P_i(x)$ for extracting the
    free coefficient, which in this case is the weighted sum of all
    messages $\sum_{j \in N_i} a_{ij} m_{ji}$.
\end{itemize}

\ignore{ Regarding the polynomial evaluation and interpolation in
steps [S2,S7] we chose the efficient implementation of
FFTEasy~\cite{FFTEasy} which uses an iterative FFT algorithm. In
this case, the polynomial is evaluated at the points \[ x_{il} =
\exp(-\frac{2 \pi j\footnote{We note $j$ as the root of the
unity}}{k} il). \] The running time of this algorithm is $d \log
d$ for both evaluation and interpolation.}

Note that the message $m_{ji}$ sent by node $j$ remains hidden
if less than  $d_i$ neighbors of $i$ collude to learn
it (this is ensured since these neighbors learn strictly less than $d_i$ values
of a polynomial of degree $d_i-1$).
The protocol requires each node $j$ to send messages to
all other
neighbors of each of its neighbors. 
We discuss the applicability of this requirement in
Section~\ref{Conclusion}.

\ignore{
\paragraph{Weighted sum} Assume that node $i$ needs to compute
the weighted sum  $\sum_{j\in N_i} a_{ij} m_{ji}$, where the
$a_{ij}$ values are constants which are known to nodes $i$ and
$j$. Then in step S4 of the protocol, node $l$ computes the sum
 $S_{li} = \sum_{j \in N_i} a_j P_{ji}(x_{il})$. The rest of the
 protocol remains as before.
}

\ignore{
Regarding the number of messages sent, we have the same overhead
as in the homomorphic encryption scheme. Note, that the size of
the messages is not increased.
}

\begin{figure}
  \includegraphics[width=250pt]{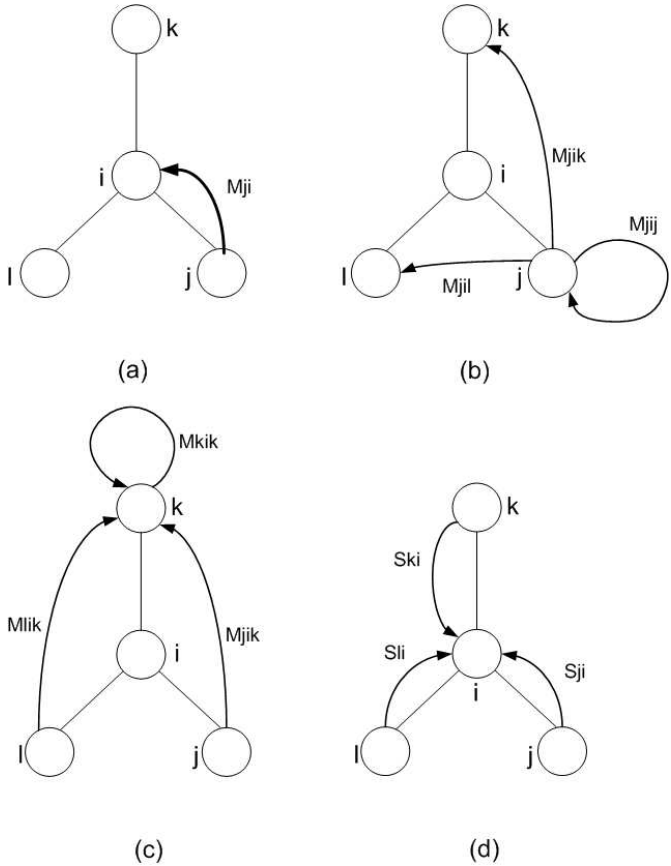}\\
  \caption{Schematic message flow in the proposed methods. The task of node $i$ is to compute the sum of all messages: $m_{ki} + m_{ji} + m_{li}$ (a) describes a message sent from $j$ to $i$ using random perturbation.
  (b) describes steps [S3] in our SSS scheme, where the same message $m_{ji}$ is split into shares sent to all of $i$ neighbors.
  (c) describes steps [S4] in our SSS scheme, where shares destined to $i$ are aggregated by its neighbors. (d) shows steps [H6] in our SSS scheme,
    which is equivalent (in term of message flow) to step [H2] in our homomorphic scheme. }\label{dimes}
\end{figure}

\subsection{Extending the method to support multiplication}
\ignore{
In the previous subsections, for simplicity of explanation, we
have resorted to the description of algorithms which perform
additions of messages in each round. Both the homomorphic and the
secret sharing scheme can be extended to support multiplication as
well.
}

Assume  that node $i$ needs to compute the multiplication of the
values of  two messages that it receives from nodes $j$ and $j'$.
\ignore{ Using the Paillier encryption, we use the homomorphic
property of exponentiation, to compute the product of two
encrypted messages. $E_{pk}(m1) \cdot_{pk} E_{pk}(m2) = E_{pk}(m1
\cdot m2)$, where the $\cdot_{pk}$ operation in the Paillier
cryptosystem is exponentiation. } The Shamir secret sharing scheme
can be extended to support multiplication using the construction
of Ben-Or, Goldwasser and Wigderson, whose details appear
in~\cite{BGW}. This requires two changes to the basic protocol.
First, the degree of the polynomials must be strictly less than
$|N_i|/2$, where $|N_i|$ is the number of neighbors of the node receiving
the messages. (This means, in particular, that security is now
only guaranteed as long as less than half of the neighbors
collude.) In addition, the neighboring nodes must exchange a
single round of messages after receiving the messages from nodes
$j$ and $j'$.
We have not implemented this variant of the protocol.

\subsection{Working in different fields}
The operations that can be applied to secrets in the Shamir secret
sharing scheme, or to encrypted values in a homomorphic encryption
scheme, are defined in a finite field or ring over which the schemes
are defined (for example, in the secret sharing case, over a field
$Z_p$ where $p$ is a prime number). The operations that we want to
compute, however, might be defined over the Real numbers.  Working
in a field is sufficient for computing additions or
multiplications of integers, if we know that the size of the field
is larger than the maximum result of the operation. If the basic
elements we work with are Real numbers, we can round them first to
the next integer, or, alternatively, first multiply them by some
constant $c$ (say, $c=10^6$) and then round the result to the
closest integer. (This essentially means that we work with
accuracy of $1/c$ if the computation involves only additions, or
an accuracy of $1/c^d$ if the computation involves summands
composed of up to $d$ multiplications.)

Handling division is much harder, since  we are essentially limited
to working with integer numbers. One possible workaround is
possible if we know in advance that a number $x$ might have to be
divided by a different number from a set $D$ (say, the numbers in
the range $[1,100]$). In that case we first multiply $x$ by the
least common multiple (lcm) of the numbers in $D$. This initial
step ensures that dividing the result by a number from $D$ results
in an integer number.

\section{Case Study: neighborhood based collaborative filtering}
\label{jacobi} To demonstrate the usefulness of our approach, we
give a specific instance of a problem our framework can solve,
preserving users' privacy. Our chosen example is in the field of
collaborative filtering. We have chosen to implement the
neighborhood based collaborative filtering algorithm, a
state-of-the-art algorithm, winner of the Netflix progress prize
of 2007. When adapting this algorithm to a Peer-to-Peer network,
there are two main challenges: first, the algorithm is
centralized, while we would like to distribute it, without losing
accuracy of the computed result. Second, we would like to add a
privacy preserving layer, which prevents the computing nodes from
learning any information about neighboring nodes or other nodes
rating, except of the computed solution.

We first describe the centralized version, and later we extend it
to be computed in a Peer-to-Peer network. Given a possibly sparse
user ratings matrix $\mR_{m \times n}$, where $m$ is the number of
users and $n$ is the number of items, each user likes to compute
an output ratings for all the items.

In the neighborhood based approach~\cite{KorenCF}, the output
rating is computed using a weighted average of the neighboring
peers:
\[ r_{ui} = \sum_{j \in N_i} r_{uj} w_{uj}. \]
Our goal is to find the weights matrix $\mW$ where $w_{ij}$
signifies the weight node $i$ assigns node $j$.

We define the following least square minimization problem for user
$i$ :
\[ \min_\vw \sum_{v \ne u}(r_{vi} - \sum_{j \in N_i}
w_{ij}r_{vj})^2\;. \]

The optimal solution is formed by differentiation and solution of
a linear systems of equations $\mR \vw = \vb$. The optimal weights
(for each user) are given by: \BE \label{eqw} \vw = (\mR^T
\mR)^{-1} \mR^T \vb \EE

We would like to distribute the neighborhood based collaborative
filtering problem to be computed in a Peer-to-Peer network. Each
peer has its own rating as input (the matching row of the matrix
$\mR$) and the goal is to compute locally, using interaction with
neighboring nodes, the weight matrix $\mW$, where each node has
the matching row in this matrix. Furthermore, the peers would like
to keep their input rating private, where no information is leaked
during the computation to neighboring or other nodes. The peers
will obtain only their matching output rating as a result of this
computation.

We propose a secure multi-party computation framework, to solve the
collaborative filtering problem efficiently and distributively,
preserving users' privacy. The computation does not reveal any
information about users' prior ratings, nor on the computed
results.

\subsection{The Jacobi algorithm for solving systems of
linear equations}\label{Jacobi} In this section we give an example
of one of the simplest iterative algorithms for solving systems of
linear equations, the Jacobi algorithm. This will serve as an
example for an algorithm our framework is able to compute, for
solving the neighborhood based collaborative filtering problem.
Note that there are numerous numerical methods we can compute
securely using our framework, among them Gauss Seidel, EM
(expectation minimization), Conjugate gradient, gradient descent,
Belief Propagation, Cholskey decomposition, principal component
analysis, SVD etc.

Given a system of linear equations $\mA \vx = \vb$, where $\mA$ is
a matrix of size $n \times n$, $\forall_i a_{ii} \ne 0$ and $\vb
\in \mathbb{R}^n$, the Jacobi
algorithm~\cite{BibDB:BookBertsekasTsitsiklis} starts from an
initial guess $\vx^{0}$, and iterates: \BE \label{Jeq}
 x_i^{r} = \frac{b_i - \sum_{j \in N_i}a_{ij} x_j^{r-1}}{ a_{ii}}
 \EE
The Jacobi algorithm is easily distributed since initially each
node selects an initial guess $x^{0}_i$, and the values $x_j^r$
are sent among neighbors. A sufficient condition for the algorithm
convergence is when the spectral radius $\rho(I - D^{-1}\mA) < 1$,
where $I$ is the identity matrix and $D = \mbox{diag}(\mA)$. This
algorithm is known to work in asynchronous settings as well. In
practice, when converging, the Jacobi algorithm convergence speed
is logarithmic in $n$\footnote{Computing the pseudo inverse
solution (equation 2) iteratively can be done more efficiently
using newer algorithms, for example ~\cite{ISIT2}. For the purpose
of the clarify of explanation, we use the Jacobi algorithm. }.

 Our goal is to compute a {\em privacy-preserving}
version of the Jacobi algorithm, where the inputs of the nodes are
private, and no information is leaked during the rounds of the
computation.

Note, that the Jacobi algorithm serves as an excellent example
since its simple update rule contains all the basic operation we
would like to support: addition, multiplication and substraction.
Our framework supports all of those numerical operations, thus
capturing numerous numerical algorithms.

\subsection{Using the Jacobi algorithm for solving the
neighborhood based collaborative filtering problem} First, we
perform a distributed preconditioning of the matrix $\mR$. Each
node $i$ divides its input row of the matrix $\mR$ by $R_{ii}$.
This simple operation is done to avoid the division in~\ref{Jeq},
while not affecting the solution vector $\vw$.

Second, since Jacobi algorithm's input is a square $n \times n$
matrix, and our rating matrix $\mR$ is of size $m \times n$, we
use the following ``trick'': We construct a new symmetric data
matrix $\tilde{\mR}$ based on the non-rectangular rating matrix
$\mR\in\mathbb{R}^{m\times n}$ \BE \label{newR}
\tilde{\mR}\triangleq\left(
  \begin{array}{cc}
    \mI_{m} & \mR^T \\
    \mR & 0 \\
  \end{array}
\right)\in\mathbb{R}^{(m+n)\times(m+n)}. \EE Additionally, we
define a new vector of variables
$\tilde{\vw}\triangleq\{\hat{\vw}^{T},\vz^{T}\}^{T}\in\mathbb{R}^{(m+n)\times1}$,
where $\hat{\vx}\in\mathbb{R}^{m\times1}$ is the (to be shown)
solution vector and $\vz\in\mathbb{R}^{n\times1}$ is an auxiliary
hidden vector, and a new observation vector
$\tilde{\vb}\triangleq\{\mathbf{0}^{T},\vb^{T}\}^{T}\in\mathbb{R}^{(m+n)\times1}$.

Now, we would like to show that solving the symmetric linear
system $\tilde{\mR}\tilde{\vw}=\tilde{\vb}$, taking the first $m$
entries of the corresponding solution vector $\tilde{\vw}$ is
equivalent to solving the original system $\mR\vw=\vb$. Note that
in the new construction the matrix $\tilde{\mR}$ is still sparse,
and has at most $2mn$ off-diagonal nonzero elements. Thus, when
running the Jacobi algorithm we have at most $2mn$ messages per
round.

Writing explicitly the symmetric linear system's equations, we get
\[ \hat{\vw}+\mR^T\vz=\mathbf{0},\mbox{  }\\
    \mR\hat{\vw}=\vb.
    \]

By extracting $\hat{\vw}$ we have \[
\hat{\vw}=(\mR^{T}\mR)^{-1}\mR^{T}\vb. \] the desired solution of
equation~\ref{eqw}.

\section{Experimental Results}
\label{exp} We have implemented our proposed framework using a
large scale simulation. Our simulation is written in C, consists
of about 1500 lines of code, and uses MPI, for running the
simulation in parallel. We run the simulation on a cluster of
Linux Pentium IV computers, 2.4Ghz, with 4GB RAM memory. We use
the open source Paillier implementation of~\cite{PaillierIMP}.

We use several large topologies for demonstrating the
applicability of our approach. The DIMES dataset~\cite{DIMES} is
an Internet router topology of around 300,000 routers and 2.2
million communication links connecting them, captured in January
2007. The Blog network, is a social network, web crawl of Internet
blogs of half a million blog sites and eleven million links
connecting them. Finally, the Netflix~\cite{Netflix} movie ratings
data, consists of around 500,000 users and 100,000,000 movie
ratings. This last topology is a bipartite graph with users at one
side, and movies at the other. This topology is not a Peer-to-Peer
network, but relevant for the collaborative filtering problem. We
have artificially created a Peer-to-Peer network, where each user
is a node, the movies are nodes as well, and edges are the ratings
assigned to the movies.

\begin{table}[h!]
\begin{center}
\begin{tabular}{|c|c|c|c|}
  \hline
  Topology & Nodes & Edges & Data Source \\
  \hline
  Blogs Web Crawl & 1.5M & 8M & IBM \\
  DIMES & 337,326  & 2,249,832 & DIMES  \\
  Netflix & 497,759 & 100M & Netflix \\
  \hline
\end{tabular}
\caption{\mbox{              } Topologies used for
experimentation}
\end{center}
\end{table}
\vspace{-5mm}

We ignore algorithm accuracy since this problem was addressed in
detail in~\cite{KorenCF}. We are mainly concerned with the
overheads of the privacy preserving mechanisms. Based on the
experimental results shown below, we conclude that the main
overhead in implementing our proposed mechanisms is the
computational overhead, since the communication latency exists
anyway in the underlying topology, and we compare the run of
algorithms with and without the added privacy mechanisms overhead.
For that purpose, we ignore the communication latency in our
simulations. This can be justified, because in the random
perturbations and homomorphic encryption schemes, we do not change
the number of communication rounds, so the communication latency
remains the same with or without the added privacy preserving
mechanisms. In the SSS scheme, we double the number of
communication rounds, so the incurred latency is doubled as well.

Table 2 compares the running times of the basic operations in the
three schemes. Each operation was repeated 100,000 times and an
average is given. As expected the heaviest computation is the
Paillier asymmetric encryption, with a security parameter of 2,048
bits. It can be easily verified, that while the SSS basic
operation takes around tens of microseconds, the Paillier basic
operations takes fractions of seconds (except of the homomorphic
multiplication which is quite efficient since it does not involve
exponentiation). In a Peer-to-Peer network, when a peer has likely
tens of connections, sending encrypted message to all of them will
take several seconds. Furthermore, this time estimation assumes
that the values sent by the function are scalars. In the vector
case, the operation will be much slower.

Table 3 outlines the running time needed to run 8 iterations of
the Jacobi algorithm, on the different topologies. Four modes of
operations are listed: no privacy preserving means we run the
algorithm without adding any privacy layer for baseline timing
comparison. Next, our three proposed schemes are shown.

In the Netflix dataset, we had to use eight computing nodes in
parallel, because our simulation memory requirement could not fit
into one processor.

As clearly shown in Table 3, our SSS scheme has significantly
reduced computation overhead relative to the homomorphic
encryption scheme, while having an equivalent level of security
(assuming that the Paillier encryption is semantically secure). In
a Peer-to-Peer network, with tens of neighbors, the homomorphic
encryption scheme incurs a high overhead on the computing nodes.

\begin{figure}
  \includegraphics[width=250pt]{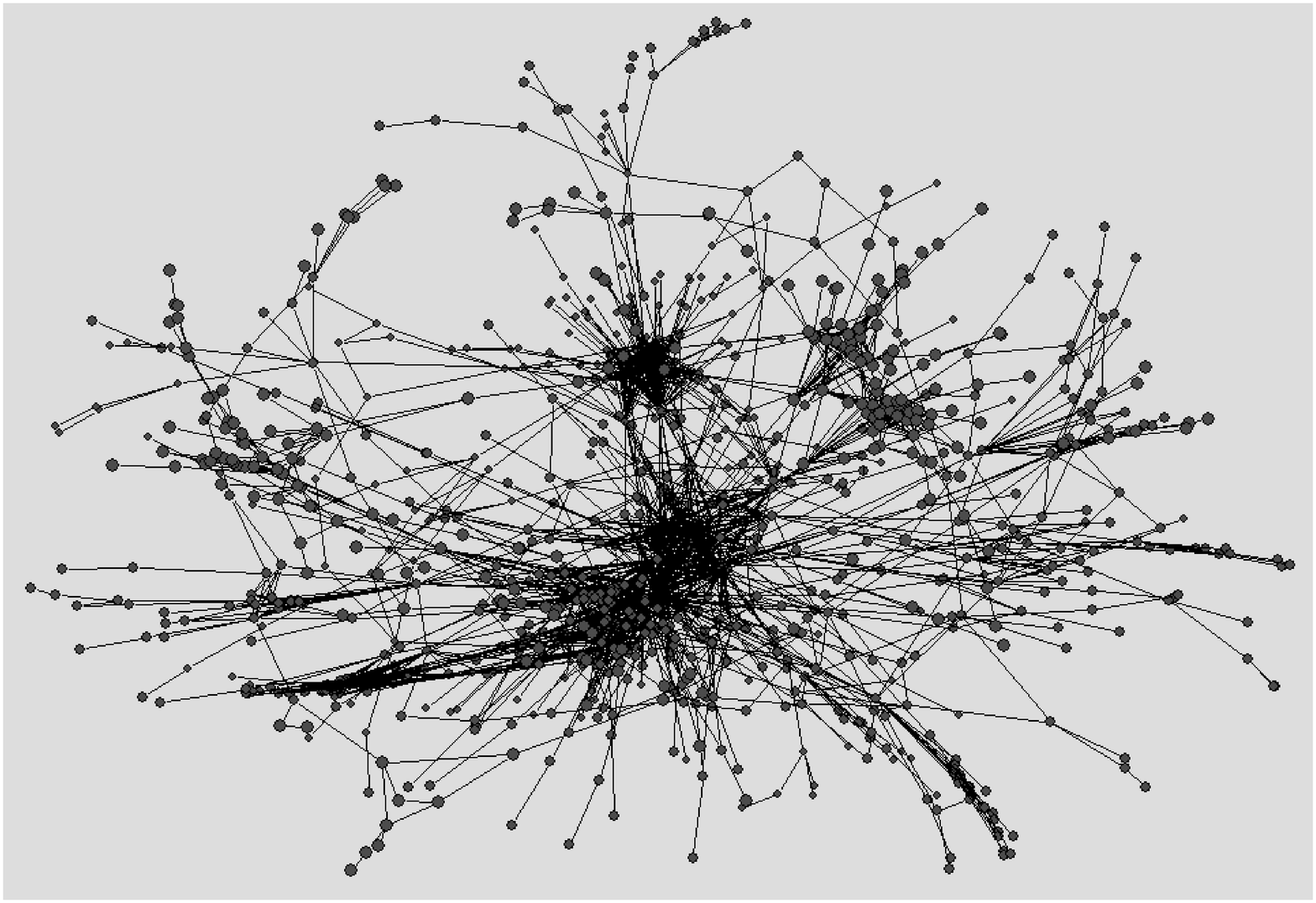}\\
  \caption{DIMES Internet router topology consisting around 300K routers and 2.2M communication links. A subgraph containing 500 nodes is shown.}\label{dimes}
\end{figure}

\begin{table*}[t!]
  \centering
\begin{tabular}{|l|l|c|c|}
  \hline
  Scheme & Operation & Time (micro second) & Msg size (bytes)\\ \hline
  Random perturbation & Adding noise & 0.0783745  & 8\\
                     & Receiver operation & $-$ &\\ \hline
  SSS & Polynomial generation and evaluation & 11.18382125 & 8 \\
          & Polynomial extrapolation & 6.13709025 & \\ \hline
  Paillier & Key generation & 5016199.4 & 2048\\
      & Encryption & 203478.62  &\\
 & Decryption & 193537.97   &\\
        & Multiplication & 99.063958 & \\
   \hline
\end{tabular}
  \label{local}
  \caption{Running time of local operations. As expected, the Paillier cryptosystem basic operations
  are time consuming relative to the SSS scheme.}\label{rt}
\end{table*}

\begin{table*}
  \centering
\begin{tabular}{|l|l|c|c|}
  \hline
  Topology & Scheme & Time (HH:MM:SS) & computing nodes\\ \hline
  DIMES & None & 0:33.36 & 1\\
  & Random Perturbations & 0:35.27 & 1\\
         & SSS & 10:53.44 & 1\\
          & Paillier & 28:44:24.00 & 1\\ \hline
  Blogs & None & 1:28.16 & 1\\
        & Random Perturbations & 1:34.85 & 1\\
         & SSS & 38:00.24 & 1\\
          & Paillier & 101:52:00.00 & 1\\ \hline
  Netflix & None & 5:31.14 & 8\\
  & Random Perturbations & 5:54.69 & 8\\
         & SSS & 21:40.00 & 8 \\
          & Paillier & - & -\\
   \hline
\end{tabular}
  \caption{Running time of eight iterations of the Jacobi algorithm. The baseline timing is compared to running without
  any privacy preserving mechanisms added. Empirical results show that computation time of the homomorphic scheme is a factor of about 1,350 times slower
  then the SSS scheme.}\label{rt}
\end{table*}

\section{Conclusion and Future Work}
\label{Conclusion} As is demonstrated by the experimental results
section, we have shown that the secret sharing scheme has the lowest computation
overhead relative to the other schemes. Furthermore, this scheme does not
involve a trusted third party, as needed by the homomorphic
encryption scheme for the threshold key generation phase. The
size of the messages sent using this method is about the same as in
the original method,
unlike the homomorphic encryption which significantly increases
message sizes. However, the  drawback of this scheme is that
neighboring nodes to node $i$ need to communicate directly between
themselves (and each message sent to node $i$ needs to be converted to
messages sent to all its neighbors). In Peer-to-Peer systems with locality property it
might be reasonable to assume that communication between the neighbors
of node $i$ is possible.
(There is a way to circumvent this requirement, by adding
asymmetric encryption. Each node will have a public key, where
message destined to this node are encrypted using its public key.
That way if node $j$ needs to send a message to node $l$, it can ask
node $i$ do deliver it, while ensuring that node $i$ does not learn
the  content of the message.
\ignore{
However, this method requires
digital signatures as well, such that node $i$ will not be able to
forge message originating from its neighbors.
}
We identify this
extension to our scheme as an area for future work.

Another area of future work is the extension of our work to
support malicious participants. The threshold Paillier
cryptosystem supports verification keys~\cite{Paillier2}, that
enable participants  to verify validity of encrypted messages.
Similarly, verifiable secret sharing schemes
like~\cite{ProactiveSS} can be used to secure secret sharing
against malicious participants, by verifying validity of
polynomial shares.

Regarding the operation in synchronous communication rounds, we
have assumed,  in order to simplify our exposition, that the
iterations of the peers are synchronized.
However, in practice it is not  valid to assume  that the clocks and message delays are synchronized in
a large Peer-to-Peer network. Luckily, it is known that linear
iterative algorithms such as the Jacobi algorithm
converge in asynchronous settings as well (meaning  that some
peers might have made more iterations than other peers but the
resulting computation will still converge to the same optimal
solution).

\bibliographystyle{latex8}
\bibliography{camera-ready-smpc-p2p08}

\begin{thebibliography}{10}\setlength{\itemsep}{-1ex}\small

\bibitem{GMP}
The {GNU MP Bignum} library. {\tt http://gmplib.org}.

\bibitem{Netflix}
{Netflix. } {\tt www.netflix.org }.

\bibitem{PaillierIMP}
Paillier {C} implementation by {John Bethencourt}. {\tt
  http://acsc.csl.sri.com/libpaillier/}.

\bibitem{AS}
R.~Agrawal and R.~Srikant.
\newblock Privacy-preserving data mining.
\newblock In {\em Proceedings of the 2000 ACM SIGMOD International Conference
  on Management of Data, May 16-18, 2000, Dallas, Texas, USA}, pages 439--450.
  ACM, 2000.

\bibitem{EWSN08}
T.~Anker, D.~Bickson, D.~Dolev, and B.~Hod.
\newblock Efficient clustering for improving network performance in wireless
  sensor networks.
\newblock {\em In European Conference on Wireless Sensor Networks (EWSN'08)}.

\bibitem{KorenCF}
R.~M. Bell and Y.~Koren.
\newblock Scalable collaborative filtering with jointly derived neighborhood
  interpolation weights.
\newblock In {\em IEEE International Conference on Data Mining (ICDM'07)},
  2007.

\bibitem{FairPlayMP}
A.~Ben-David, N.~Nisan, and B.~Pinkas.
\newblock Fairplaymp -- a system for secure multi-party computation.
\newblock manuscript, 2008.

\bibitem{BGW}
M.~Ben-Or, S.~Goldwasser, and A.~Wigderson.
\newblock Completeness theorems for non-cryptographic fault-tolerant
  distributed computation.
\newblock In {\em 20th STOC, 1988, pp. 1-10.}

\bibitem{BibDB:BookBertsekasTsitsiklis}
D.~P. Bertsekas and J.~N. Tsitsiklis.
\newblock {\em Parallel and Distributed Calculation. Numerical Methods.}
\newblock Prentice Hall, 1989.

\bibitem{p2p-rating}
D.~Bickson, D.~Malkhi, and L.~Zhou.
\newblock Peer to peer rating.
\newblock {\em In the 7th IEEE Peer-to-Peer Computing, Galway, Ireland, Sept.
  2007.}

\bibitem{ISIT2}
D.~Bickson, O.~Shental, P.~H. Siegel, J.~K. Wolf, and D.~Dolev.
\newblock Gaussian belief propagation based multiuser detection.
\newblock In {\em IEEE Int. Symp. on Inform. Theory (ISIT), Toronto, Canada,
  July 2008, to appear.}

\bibitem{Canny}
J.~Canny.
\newblock Collaborative filtering with privacy via factor analysis.
\newblock In {\em SIGIR '02: Proceedings of the 25th annual international ACM
  SIGIR conference on Research and development in information retrieval}, pages
  238--245, New York, NY, USA, 2002. ACM.

\bibitem{DiNi}
I.~Dinur and K.~Nissim.
\newblock Revealing information while preserving privacy.
\newblock In {\em PODS '03: Proceedings of the twenty-second ACM
  SIGMOD-SIGACT-SIGART symposium on Principles of database systems}, pages
  202--210, New York, NY, USA, 2003. ACM.

\bibitem{RandomNoise}
H.~Dutta, H.~Kargupta, S.~Datta, and K.~Sivakumar.
\newblock Analysis of privacy preserving random perturbation techniques:
  further explorations.
\newblock In {\em WPES '03: Proceedings of the 2003 ACM workshop on Privacy in
  the electronic society}, pages 31--38, New York, NY, USA, 2003. ACM.

\bibitem{Paillier2}
P.-A. Fouque, G.~Poupard, and J.~Stern.
\newblock Sharing decryption in the context of voting or lotteries.
\newblock In {\em Financial Cryptography, volume 1962 of Lecture Notes in
  Computer Science, pages 90–104. Springer, 2001}.

\bibitem{GMW}
O.~Goldreich, S.~Micali, and A.~Wigderson.
\newblock How to play any mental game or {A} completeness theorem for protocols
  with honest majority.
\newblock In {\em Proceedings of the 19th Annual Symposium on Theory of
  Computing ({STOC})}, pages 218--229, New York, NY USA, May 1987. ACM Press.

\bibitem{ProactiveSS}
A.~Herzberg, S.~Jarecki, H.~Krawczyk, and M.~Yung.
\newblock Proactive secret sharing, or: How to cope with perpetual leakage.
\newblock In {\em Advances in Cryptology---CRYPTO '95, volume 963 of Lecture
  Notes in Computer Science, pages 339--352, Berlin, 1995. Springer-Verlag.}

\bibitem{EigenTrust}
S.~D. Kamvar, M.~T. Schlosser, and H.~G. Molina.
\newblock The eigentrust algorithm for reputation management in p2p networks.
\newblock In {\em Proceedings of the Twelfth International World Wide Web
  Conference, 2003.}

\bibitem{Fairplay}
D.~Malkhi, N.~Nisan, B.~Pinkas, and Y.~Sella.
\newblock Fairplay --- a secure two-party computation system.
\newblock In {\em Proc. Usenix Security Symposium 2004}, 2004.

\bibitem{Separable}
D.~Mosk-Aoyama and D.~Shah.
\newblock Computing separable functions via gossip.
\newblock In {\em PODC '06: Proceedings of the twenty-fifth annual ACM
  symposium on Principles of distributed computing}, pages 113--122, New York,
  NY, USA, 2006. ACM Press.

\bibitem{Paillier}
P.~Paillier.
\newblock Public-key cryptosystems based on composite degree residuosity
  classes.
\newblock In {\em EUROCRYPT '99, Springer-Verlag (LNCS 1592)}.

\bibitem{BibDB:BookPearl}
J.~Pearl.
\newblock {\em Probabilistic Reasoning in Intelligent Systems: Networks of
  Plausible Inference}.
\newblock Morgan Kaufmann, San Francisco, 1988.

\bibitem{SSS}
A.~Shamir.
\newblock "how to share a secret".
\newblock In {\em {Communications of the ACM, 22(1), pp 612–613, 1979}}.

\bibitem{DIMES}
Y.~Shavitt and E.~Shir.
\newblock Dimes: Let the internet measure itself.
\newblock {\em {ACM} SIGCOMM Computer Communications Review}, 35(5):71--74,
  2005.

\bibitem{PP3}
L.~Wan, W.~K. Ng, S.~Han, and V.~C.~S. Lee.
\newblock Privacy-preservation for gradient descent methods.
\newblock In {\em KDD '07: Proceedings of the 13th ACM SIGKDD international
  conference on Knowledge discovery and data mining}, pages 775--783, New York,
  NY, USA, 2007. ACM.

\bibitem{Yao}
A.~Yao.
\newblock Protocols for secure computations.
\newblock In {\em Proceedings of the 23rd Symposium on Foundations of Computer
  Science ({FOCS})}, pages 160--164. {IEEE} Computer Society Press, 1982.

\bibitem{PP2}
S.~Zhang, J.~Ford, and F.~Makedon.
\newblock A privacy-preserving collaborative filtering scheme with two-way
  communication.
\newblock In {\em EC '06: Proceedings of the 7th ACM conference on Electronic
  commerce}, pages 316--323, New York, NY, USA, 2006. ACM.

\end{thebibliography}
\end{document}